\documentstyle[12pt]{article}
\begin{document}

\begin{center}
{\bf The averaging of Hamiltonian structures in discrete
variant of Whitham method}
\end{center}

\begin{center}
A.Ya.Maltsev
\end{center}

\begin{center}
L.D.Landau Institute for Theoretical Physics,
e-mail: maltsev@itp.ac.ru
\end{center}

In this work we shall construct the 
procedure of averaging of Hamiltonian structures
in discrete variant of Whitham method.
This procedure will be analogous to Dubrovin-Novikov procedure
of averaging of local field-theoretical brackets.
In discrete variant of Whitham method
we obtain the evolution of slow modulated m-phase
solutions of nonlinear systems of type:
\begin{equation}
\label{dsyst}
\dot{\varphi}^{i}_{n} = f^{i}(\mbox{$\boldmath \varphi$}_{n},
\mbox{$\boldmath \varphi$}_{n-1},
\mbox{$\boldmath \varphi$}_{n+1},\dots) ,
\end{equation}
defined on the space of fields
$\mbox{$\boldmath \varphi$} =
(\varphi^{1}_{n},\dots,\varphi^{p}_{n})$, where
$n$ runs all integer numbers and functions
$f$ depend on finite number of arguments.

Quasi-periodic m-phase solutions of system (\ref{dsyst})
are the solutions of the form:
\begin{equation}
\label{disresh}
\varphi^{i}_{n}(t) = \Phi^{i}({\bf k}({\bf U})n +
\mbox{$\boldmath \omega$}({\bf U})t + \theta_{0},{\bf U}) ,
\end{equation}
where ${\bf k}({\bf U})$ and
$\mbox{$\boldmath \omega$}({\bf U})$ are m-dimensional vector,
${\bf U} = (U^{1},\dots,U^{N})$ are parameters of solution.
Here $2\pi$-periodic functions of m variables $(\theta)$
$\Phi^{i}(\theta,{\bf U})$ satisfy to the conditions:
\begin{equation}
\label{dispod}
\omega^{\alpha}({\bf U}) \Phi^{i}_{\theta^{\alpha}}(\theta,{\bf U}) =
f^{i}(\mbox{$\boldmath \Phi$}(\theta,{\bf U}),
\mbox{$\boldmath \Phi$}(\theta-{\bf k}({\bf U}),{\bf U}),
\mbox{$\boldmath \Phi$}(\theta+{\bf k}({\bf U}),{\bf U}),
\dots) .
\end{equation}

Let system (\ref{dsyst}) is Hamiltonian with respect to
Poisson bracket:
\begin{equation}
\label{disskob}
\{\varphi^{i}_{n},\varphi^{j}_{k}\} =
\sum_{m} B^{ij}_{(m)}
(\mbox{$\boldmath \varphi$}_{n},
\mbox{$\boldmath \varphi$}_{n-1},
\mbox{$\boldmath \varphi$}_{n+1},\dots) \,\,\,
\delta_{n, k-m} ,
\end{equation}
where the sum contains a finite number of terms (the description of
such bracket can be found in [3]), with Hamiltonian function:

\begin{equation}
\label{disgam}
H [\mbox{$\boldmath \varphi$}] =
\sum_{m=-\infty}^{\infty} {\cal P}_{H}
(\mbox{$\boldmath \varphi$}_{n},
\mbox{$\boldmath \varphi$}_{n-1},
\mbox{$\boldmath \varphi$}_{n+1},\dots)
\end{equation}
and has $N$ commutative conservation laws of the form:
\begin{equation}
\label{disint}
I^{\nu} = \sum_{n=-\infty}^{\infty}
{\cal P}^{\nu}
(\mbox{$\boldmath \varphi$}_{n},
\mbox{$\boldmath \varphi$}_{n-1},
\mbox{$\boldmath \varphi$}_{n+1},\dots) , \,\,\,\,\,
\nu = 1,\dots,N ,
\end{equation}
\begin{equation}
\label{commut}
\{I^{\nu} , I^{\mu}\} = 0 , \,\,\,\,\,
\{I^{\nu} , H\} = 0
\end{equation}
($B^{ij}_{(m)}$, ${\cal P}_{H}$ and ${\cal P}^{\nu}$
depend on finite number of arguments).

We shall have view (\ref{dsyst}):
$${d \over dt} {\cal P}^{\nu}
(\mbox{$\boldmath \varphi$}_{n},
\mbox{$\boldmath \varphi$}_{n-1},\dots) =
\sum_{m} [ Q^{\nu}_{(m)}
(\mbox{$\boldmath \varphi$}_{n},
\mbox{$\boldmath \varphi$}_{n-1},\dots) -
Q^{\nu}_{(m)}
(\mbox{$\boldmath \varphi$}_{n-m},
\mbox{$\boldmath \varphi$}_{n-m-1},\dots)]$$
(there is a finite number of terms in the sum,
$Q^{\nu}_{(m)}$ depend on finite number of arguments).
Whitham system, giving the evolution of slow modulated parameters
${\bf U}$, can be written in the form:
\begin{equation}
\label{conserv}
\partial/\partial T \,\, \langle {\cal P}^{\nu}\rangle ({\bf U}(X)) =
\sum_{m} m \,\,\, \partial/\partial X \,\,
\langle Q^{\nu}_{(m)}\rangle ({\bf U}(X)) ,
\end{equation}
here $\langle\dots\rangle$ means the averaging on the family
(\ref{disresh}), defined by the formula:
$$\langle F
(\mbox{$\boldmath \varphi$}_{n},
\mbox{$\boldmath \varphi$}_{n-1},\dots)
\rangle ({\bf U}) =
1/(2\pi)^{m} \int F(\mbox{$\boldmath \Phi$}(\theta,{\bf U}),
\mbox{$\boldmath \Phi$}(\theta-{\bf k}({\bf U}),{\bf U}),
\dots) d^{m}\theta .$$

View (\ref{disskob}) we shall have:
\begin{equation}
\label{skobint}
\{{\cal P}^{\nu}(\mbox{$\boldmath \varphi$}_{n},\dots),
{\cal P}^{\mu}(\mbox{$\boldmath \varphi$}_{k},\dots)\} =
\sum_{m} A^{\nu\mu}_{(m)}
(\mbox{$\boldmath \varphi$}_{n},
\mbox{$\boldmath \varphi$}_{n-1},\dots)
\,\,\, \delta_{n, k-m}
\end{equation}
and, view
(\ref{commut}),
$$\sum_{m} A^{\nu\mu}_{(m)}
(\mbox{$\boldmath \varphi$}_{n},
\mbox{$\boldmath \varphi$}_{n-1},\dots) =
\sum_{l} \left( Q^{\nu\mu}_{(l)}
(\mbox{$\boldmath \varphi$}_{n},
\mbox{$\boldmath \varphi$}_{n-1},\dots) -
Q^{\nu\mu}_{(l)}
(\mbox{$\boldmath \varphi$}_{n-l},
\mbox{$\boldmath \varphi$}_{n-l-1},\dots) \right)$$
(all sums contain finite numbers of terms,
$A^{\nu\mu}_{(m)}$ and $Q^{\nu\mu}_{(l)}$ depend on
finite numbers of arguments).

Theorem 1. Let parameters
${\bf U}$ are such, that
$U^{\nu} = \langle{\cal P}^{\nu}\rangle$, then:\newline
1) System (\ref{conserv}) is Hamiltonian with respect
to Poisson bracket of Hydrodynamic Type (see [4]-[5]),
defined by the relations:
\begin{equation}
\label{gdskob}
\{U^{\nu}(X),U^{\mu}(Y)\} =
\sum_{m} m \, \langle A^{\nu\mu}_{(m)}\rangle(X)
\delta^{\prime}(X-Y) +\left[{\partial \over \partial X}
\sum_{l} l \, \langle Q^{\nu\mu}_{(l)}\rangle(X)
\right] \delta(X-Y)
\end{equation}
with Hamiltonian function: $H = \int \langle {\cal P}_{H}\rangle (X) dX$.

2) From (\ref{gdskob}) follow the relations:
$$\{k^{\alpha}({\bf U}(X)), U^{\nu}(Y)\} =
\omega^{\alpha(\nu)}({\bf U}(X))\, \delta^{\prime}(X-Y) +
{\partial \omega^{\alpha(\nu)}({\bf U}(X)) \over \partial X}\,
\delta(X-Y) ,$$
\begin{equation}
\label{sootn}
\{k^{\alpha}({\bf U}(X)), k^{\beta}({\bf U}(Y))\} = 0, \,\,\,
\alpha, \beta = 1,\dots,m, \,\, \nu = 1,\dots,N ,
\end{equation}
where $\mbox{$\boldmath \omega$}^{(\nu)}({\bf U})$ are
frequencies analogous to $\mbox{$\boldmath \omega$}$ and
corresponding to flows generated by integrals
$I^{\nu}$ on the family of solutions (\ref{disresh}).

Theorem 2. Let system (\ref{dsyst}) has two different
sets of integrals of form (\ref{disint}),
$\{I^{1},\dots,I^{N}\}$ and $\{{\bar I}^{1},\dots,{\bar I}^{N}\}$,
possessing all necessary properties. Then brackets
(\ref{gdskob}) defined with the aid of these sets are coincide
with each other. That is, if on the family (\ref{disresh})
we have:
$U^{\nu} = I^{\nu}$, ${\bar U}^{\nu} = {\bar I}^{\nu}$,
then bracket (\ref{gdskob}), defined with the aid of the set
$\{I^{1},\dots,I^{N}\}$, transforms into the bracket
defined with the aid of the set
$\{{\bar I}^{1},\dots,{\bar I}^{N}\}$ after the
transformation of parameters $U^{\nu} = U^{\nu}({\bar {\bf U}})$
of m-phase solutions (\ref{disresh}).

The proof of Theorems 1 and 2 is analogous to the proof
of the same theorems for the averaged by Dubrovin-Novikov method
field-theoretical brackets (см.[6],[7]) and is based on
Dirac restriction of Poisson bracket:
$$\{\varphi^{i}(\theta,X),\varphi^{j}(\theta^{\prime},Y)\} =
\sum_{m} B^{ij}_{(m)}(\mbox{$\boldmath \varphi$}(X),
\mbox{$\boldmath \varphi$}(X-\epsilon),\dots)
\delta(X-Y+m\epsilon) \delta(\theta - \theta^{\prime})$$
on the sub-manifold of functions, such that
$\mbox{$\boldmath \varphi$}(\theta, X)$ at any $X$
is the function from (\ref{dispod}) (and we use the expansion
$\delta(X-Y+m\epsilon) = \delta(X-Y) +
m\epsilon \delta^{\prime}(X-Y) + \dots$ in the convolution
with smooth functions.)
The necessary conditions of regularity of this sub-manifold
and the coordinates used for restriction are described in [6].

The author is grateful to S.P.Novikov and O.I.Mokhov
for the attention to the work.

The work was done under the financial support of
Russian Foundation for Fundamental Research (grant
96-01-01623) and INTAS (grant INTAS 96-0770).

\centerline{References.}

$[1]$  G. Whitham // Linear and Nonlinear waves.
Wiley, New York, 1974.

$[2]$ Luke J.C.// Proc. Roy. Soc. London Ser. A, 1966.
V. 292. No. 1430, 403-412.

$[3]$  B.A.Dubrovin// Functional Analysis and It's Appl., 1989.
Vol. 23, N. 2, P. 57-59.

$[4]$ B.A.Dubrovin, S.P.Novikov //
Dokl. Akad. Nauk SSSR, 1983. Vol. 270, N. 4, P. 781-785.

$[5]$ B.A.Dubrovin, S.P.Novikov //
Uspekhi Mat. Nauk (Russian Mat. Surv.), 1989. Vol. 44, N. 6, P. 29-98.

$[6]$ A.Ya.Maltsev. //"The conservation of the Hamiltonian
structures in Whitham's method of averaging".
solv-int/9611008, 1996 (to appear) .

$[7]$ A.Ya.Maltsev. // Uspekhi Mat. Nauk. 1997. Vol. 52, N. 2, P. 177-178.
\end{document}